\documentclass[prl,letterpaper,english,reprint,nofootinbib,aps,superscriptaddress,showpacs,showkeys]{revtex4-1}

\usepackage{bm}
\usepackage{float}

\usepackage{babel,calc,amsmath,amsthm,amssymb,graphicx,subfigure,xcolor,comment}
\usepackage{mathdots}
\usepackage[T1]{fontenc}
\usepackage[unicode=true]{hyperref}
\hypersetup{
	colorlinks=true,       		
	linkcolor=red,          	
	citecolor=blue,             
	urlcolor=blue,          
}

\usepackage{booktabs}
\renewcommand{\arraystretch}{1.2}
\usepackage{makecell}
\usepackage{braket}
\usepackage[normalem]{ulem}

\newif\ifdebug

\debugtrue

\ifdebug
\definecolor{zhliu}{rgb}{0.48, 0.12, 0}

\newcommand{\note}[1]{\textcolor{zhliu}{#1}}
\newcommand\delete{\bgroup\markoverwith{\textcolor{zhliu}{\rule[0.5ex]{2pt}{0.8pt}}}\ULon}

\else

\newcommand{\note}[1]{\ignorespaces}
\newcommand{\delete}[1]{\ignorespaces}

\fi

\begin{document}
	
	\renewcommand{\figurename}{Fig.}
	
	\title{Robust Steerability Classification via Key Feature Extraction and Matrix Structure Preservation}

	\author{Yutao Xin}
	\affiliation{
		School of Mathematics and Physics,
		North China Electric Power University,
		Beijing 102206, China
	}
	\author{Huixian Meng}
	\email{ huixianmenghd@ncepu.edu.cn}
	\affiliation{
		School of Mathematics and Physics,
		North China Electric Power University,
		Beijing 102206, China
	}
	\author{Zhongyan Li}
	\affiliation{
		School of Mathematics and Physics,
		North China Electric Power University,
		Beijing 102206, China
	}
	\author{Pu Wang}
	\affiliation{
		School of Control and Computer Engineering,
		North China Electric Power University,
		Beijing 102206, China
	}
	
	\date{\today}
	\date{\today}

	\begin{abstract}
		
		Generalization ability is essential for assessing the robustness of quantum steerability classifiers. In this work, we investigate robust steerability classification from the perspective of key feature extraction and matrix structure preservation. The dataset introduced in Phys. Rev. A 100, 022314 (2019) provides the training basis for the present work. With strictly unsteerable random states, $T$-diagonal states, and All-Versus-Nothing (AVN) states, we evaluate the generalization performance of support vector machines (SVMs), multilayer perceptrons (MLPs), and deep perceptron
		 control classifiers(DPs) trained on full-information features. None of these classifiers perform consistently on $T$-diagonal or AVN states. Given that stochastic local operations and classical communication and local unitary transformations preserve steerability, we introduce a key feature that determines steerability. SVMs trained on this feature overcome the instability on $T$-diagonal states but still fail on AVN states. Moreover, this feature alone is insufficient for training robust neural-network-based steerability classifiers. Recognizing that flattening quantum states into one-dimensional vectors may destroy their intrinsic matrix structure, we introduce matrix versions of both features and train convolutional neural networks on them. The most robust overall performance among the tested classifiers is achieved only when the matrix structure is preserved and key features are extracted simultaneously. Finally, as an application, we employ the most robust classifiers to predict the number of projective measurements required to detect the steerability of axially symmetric states.
		
	\end{abstract}
	
	
	\maketitle

	\section{Introduction}
	
	Quantum steering, first introduced by Schr\"odinger in response to the Einstein-Podolsky-Rosen paradox~\cite{schrodinger1935,einstein1935}, describes the ability of one party to remotely affect the state of another party through local measurements. Since its rigorous formulation by Wiseman \textit{et al.}~\cite{wiseman2007}, steering has been recognized as a fundamental and asymmetric form of quantum correlation, lying between entanglement and Bell nonlocality~\cite{uola2020}. Quantum steering has become an important resource in quantum information for one-sided device-independent quantum key distribution~\cite{branciard2012,gehring2015}, quantum channel discrimination~\cite{piani2015}, and randomness certification~\cite{law2014}.
	
	Detecting steerability remains a challenging task. Steering inequalities provide sufficient criteria~\cite{cavalcanti2009,saunders2010,schneeloch2013}, while semidefinite programming (SDP) offers a way to test the existence of local-hidden-state models~\cite{cavalcanti2016,cavalcanti2017}. Nevertheless, SDP-based methods are computationally demanding and become costly for large datasets or more complicated measurement scenarios. In recent years, machine learning has provided data-driven tools for quantum information and many-body physics~\cite{carrasquilla2017,carleo2019,gao2018,deng2018}. In the context of quantum steering, supervised and semi-supervised learning methods have been applied to steerability detection and steering-boundary estimation~\cite{ku2018,ren2019,chenZ2021,He2022detecting,Wang2024,coppola2022,wangpu}. Recent work has emphasized the measurement-dependent nature of quantum steering by learning the hierarchy of steering measurement settings~\cite{Wang2024}, suggesting that the number of measurement settings should be treated as a physically relevant factor rather than a mere numerical parameter.

	Generalization ability is the key to judging the robustness of steerability classifiers. In this paper, we shall use the following five families of states to check the generalization of classifiers: Werner states~\cite{Werner1989}, strictly unsteerable random states~\cite{Hirsch2016,CavalcantiD2016},  two- and three-measurement test sets constructed from $T$-diagonal states~\cite{Quan2016,Nguyen2020}, All-Versus-Nothing (AVN) states~\cite{chen2013}. The classifiers include support vector machines (SVMs)~\cite{ren2019}, multilayer perceptrons (MLPs)~\cite{He2022detecting}, and deep perceptron
	control classifiers(DPs)~\cite{Wang2024}. 
	
	Since stochastic local operations and classical communication (SLOCC) performed on Bob's subsystem, together with local unitary transformations on the bipartite system, preserve steerability from $A$ to $B$, we shall extract steerability-determined feature to train robust classifiers. Existing machine learning models for detecting quantum steerability usually flatten quantum states into one-dimensional feature vectors as inputs. However, this operation may destroy the intrinsic matrix structure and steering-relevant correlations encoded in quantum states. 
	
	In this paper, we shall introduce the matrix versions of the one-dimensional full-information feature and  steerability-determined feature, and train convolutional neural networks  (CNNs). Encouragingly, the CNNs trained by matrix version of the steerability-determined feature perform best. These results indicate that both the matrix-structure-preserving and the extracting of governing steerability
	critical features enable the training of CNNs  with the most robust steerability classification capacity. As applications, we use the most robust classifiers to predict how many projective measurements (PMs) are needed to detect the steerability of steerable axially symmetric states.

	\section{PRELIMINARIES}
	
	Consider a bipartite quantum state $\rho^{AB}$ shared by Alice and Bob.
	Alice performs $m_A$  local measurements on her subsystem
	$\mathcal{M}=\{M_x=\{M_{a|x}\}_{a=1}^{o_A}\}_{x=1}^{m_A}$
	, where $x$ labels the measurement setting and
	$a\in\{1,2,\ldots,o_A\}$ denotes the corresponding outcome.
	For each pair $(a,x)$, the unnormalized conditional state prepared on Bob's subsystem is given by
	\begin{equation}
		\sigma_{a|x}
		=
		\mathrm{Tr}_A\left[
		(M_{a|x}\otimes I_B)\rho^{AB}
		\right].
	\end{equation}
	The collection $\{\sigma_{a|x}\}$ is called an assemblage.
	
	The assemblage is said to admit a local-hidden-state (LHS) model if it can be written as
	\begin{equation}
		\sigma_{a|x}
		=
		\int p(\lambda)p(a|x,\lambda)\sigma_\lambda\,d\lambda,
		\quad \forall a,x ,
		\label{eq:lhs_model}
	\end{equation}
	where $p(\lambda)$ is a hidden-variable distribution,
	$p(a|x,\lambda)$ is Alice's local response function, and
	$\{\sigma_\lambda\}$ is a set of local quantum states for Bob.
	If such a decomposition exists, the assemblage is unsteerable under the measurement set $\mathcal{M}$; otherwise, it demonstrates steering from Alice to Bob.
	In the general sense, $\rho^{AB}$ is called unsteerable if an LHS model exists for all allowed local measurements, and steerable if at least one measurement set leads to an assemblage that does not admit an LHS model~\cite{wiseman2007}.
	Unless otherwise specified, the steering direction considered in this work is from Alice to Bob.
	\subsection{Semidefinite Programming}
	
	Determining whether an assemblage admits an LHS model can be formulated as a SDP feasibility problem~\cite{cavalcanti2017}.
	For a fixed measurement set $\mathcal{M}$, the assemblage $\{\sigma_{a|x}\}$ is unsteerable if there exist positive semidefinite operators $\{\sigma_\lambda\}$ such that
	\begin{equation}
		\begin{aligned}
			\text{find} \quad & \{\sigma_\lambda\}, \quad \sigma_\lambda \geqslant 0,\\
			\text{s.t.} \quad
			& \sum_{\lambda}D(a|x,\lambda)\sigma_\lambda
			=
			\sigma_{a|x},
			\quad \forall a,x .
		\end{aligned}
		\label{eq:sdp_primal}
	\end{equation}
	Here, $D$ denotes the deterministic response function associated with the hidden variable $\lambda$, assigning a fixed outcome to each measurement setting. Hence, Eq.~\eqref{eq:sdp_primal} is feasible exactly when the given assemblage admits an LHS decomposition.
	
	The corresponding dual formulation can be interpreted in terms of steering witnesses. In particular, if one can find Hermitian operators $\{F_{a|x}\}$ satisfying
	\begin{equation}
		\sum_{a,x}D(a|x,\lambda)F_{a|x}\geqslant 0,
		\quad \forall \lambda ,
	\end{equation}
	while giving
	\begin{equation}
		\mathrm{Tr}\left(
		\sum_{a,x}F_{a|x}\sigma_{a|x}
		\right)<0,
	\end{equation}
	then the assemblage violates the LHS constraints and is therefore steerable. For a general quantum state, proving its unsteerability requires constructing a local hidden state model, which remains
	a challenging task in practice.
	
	This computational difficulty motivates the use of machine-learning methods for steerability classification.
	Many previous studies represent density matrices by one-dimensional feature vectors~\cite{ren2019,He2022detecting,Wang2024}.
	While such vectorized features are convenient for standard classifiers, they may destroy the matrix-structure information encoded in quantum states.
	In this work, we introduce matrix-structured features and train CNN to classify steerability.
	
	\subsection{Convolutional Neural Networks}
	
	CNNs are a class of deep learning architectures designed to process data with grid-like structures~\cite{carrasquilla2017}. Unlike fully connected networks, which usually operate on flattened vectors~\cite{He2022detecting,Wang2024}, CNNs use local connectivity and weight sharing to extract features while preserving the relative arrangement of input elements. This property is suitable for quantum steering classification, since two-qubit quantum states are naturally represented by complex-valued $4\times4$ density matrices.
	
	In this work, each input sample, indexed by $\ell$, is represented as a multi-channel array
	$\mathcal{X}^{(\ell)}\in\mathbb{R}^{C\times H\times W}$.
	For two-qubit states, $H=W=4$.
	Since the density matrices are complex-valued, we decompose each matrix into its real and imaginary parts and stack them as two channels, so that $C=2$.
	
	A convolutional layer applies learnable kernels $K^{(k)}\in\mathbb{R}^{C\times f\times f}$ to the input array. The pre-activation output of the $k$-th kernel at spatial position $(i,j)$ is given by
	\begin{equation}
		Y_{i,j}^{(k)}
		=
		\sum_{c=1}^{C}
		\sum_{p=0}^{f-1}
		\sum_{q=0}^{f-1}
		\mathcal{X}_{c,i+p,j+q}
		K_{c,p,q}^{(k)}
		+
		b^{(k)},
		\label{eq:convolution}
	\end{equation}
	where $b^{(k)}$ is the bias term. The output is then passed through a nonlinear activation function. In this work, we use the rectified linear unit (ReLU),
	\begin{equation}
		Z_{i,j}^{(k)}
		=
		\max\left(0,Y_{i,j}^{(k)}\right).
	\end{equation}
	After several convolutional blocks, the extracted matrix-structured features are aggregated and fed into the final classifier for steerability prediction.

	\section{Generalization capacity of SVM based on  Vectorized Feature $F_1$}

	For two-qubit quantum states, Ren \textit{et al.} introduced the full-information feature $F_1$~\cite{ren2019}, which is a $16$-dimensional real vector. For a $4\times4$ density matrix $\rho_{AB}$, the input feature $F_1$ is formed by the first three diagonal entries and the real and imaginary parts of the six sub-diagonal elements. For different numbers of measurement settings, the steerability label, the last element of $F_1$, is obtained by the SDP method~\cite{ren2019}. Unless otherwise stated, the random-state steering classification dataset employed throughout this paper is that presented in reference~\cite{ren2019}. All accuracies reported in this paper are listed in the Appendix A.

	Based on the random-state steering classification dataset,  Ren \textit{et al.} trained SVM classifiers. For the full-information feature $F_1$, they obtained high cross-validation and test accuracies, as well as good generalization performance on Werner states, which have the form
	\begin{equation}
		\rho_W(p)
		=
		p|\psi^-\rangle\langle\psi^-|
		+
		(1-p)\frac{I_4}{4},\
		|\psi^-\rangle
		=
		\frac{|01\rangle-|10\rangle}{\sqrt{2}}.
	\end{equation}
	
	In this work, we train the  $C$ and $\gamma$ parameters of SVM, and list their optimal values in  Appendix  B to  facilitate result reproduction. For random states, the cross-validation accuracies and test accuracies  exceed $95\%$ for any number of $m$ PMs. In contrast, when evaluating generalization on Werner states, we observe that all test accuracies are over $91\%$, except the case $m=3$, where the accuracy drops to approximately $83\%$.
	
\begin{figure*}[t]
		\centering
		\includegraphics[width=\textwidth]{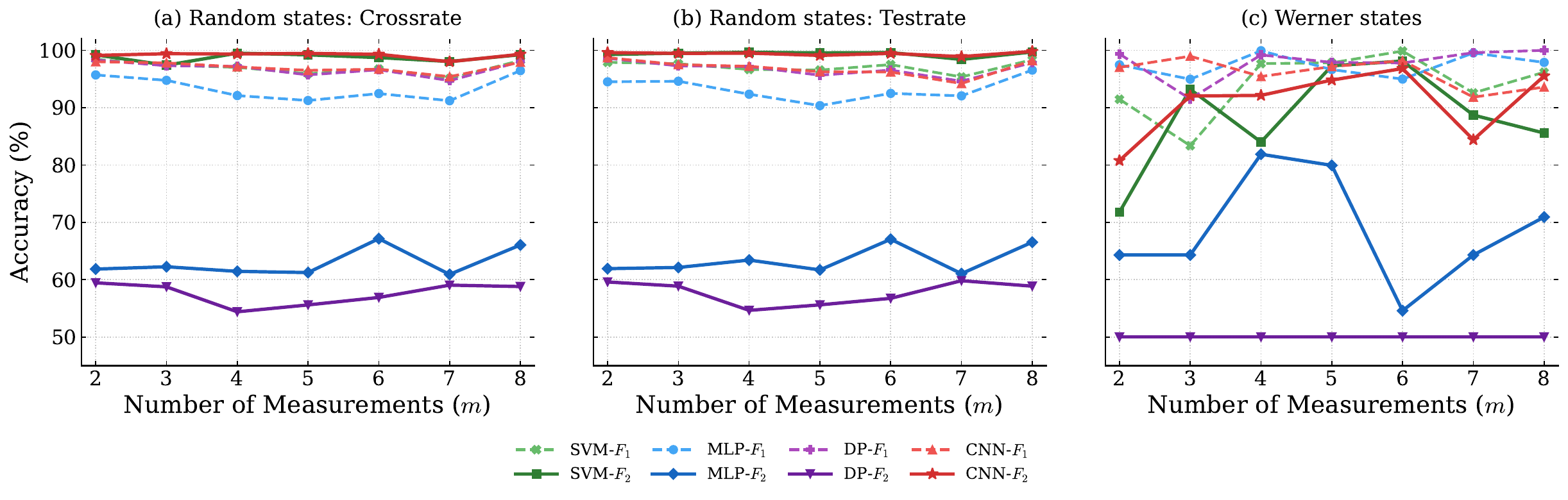}
		\caption{
			Baseline comparison of different classifiers using $F_1$ and $F_2$.
			(a) Cross-validation accuracy on random states.
			(b) Test accuracy on random states.
			(c) Prediction accuracy on Werner states.
			The detailed numerical values are listed in the Appendix~\ref{app:accuracy}.
		}
		\label{fig:random_werner_allmodels}
	\end{figure*}

	The green dashed lines in Figure~\ref{fig:random_werner_allmodels} show the cross-validation accuracies and test accuracies on random states, as well as the prediction accuracies on Werner states.
	
	The positive samples in the training set are quantum states for which SDP cannot detect steerability after randomly selecting $100$ sets of $m$ PMs. However, these states are not necessarily unsteerable under a fixed number of PMs.  In this paper, we shall use a class of strictly unsteerable random states~\cite{CavalcantiD2016} to evaluate classifiers' generalization.
	Reference~\cite{Nguyen2020} presented a sufficient and necessary condition for detecting the steerability of  $T$-diagonal states under two PMs, as well as  a sufficient condition under three. For finite and more than three PMs, no analytical prediction is available. All-Versus-Nothing (AVN) states~\cite{chen2013} are a kind of mixed states which are entangled if and only if they are steerable. Especially, if an AVN state is entangled, then it is steerable under two PMs.

	In the following, we evaluate the generalization of SVM-$F_1$ on five families of states: strictly unsteerable random states, $T$-diagonal states that are (i) unsteerable under two PMs, (ii) steerable under two PMs, and (iii) steerable under three PMs, and AVN states that are steerable under two PMs.

	\subsection{Strictly Unsteerable Random States}
	
	We first consider strictly unsteerable random states. Specifically, for a given quantum state $\rho_{AB}$, one considers the following optimization problem:
	\begin{equation}
		\begin{aligned}
			\max_{O_{AB},\sigma_{\lambda}}
			&\quad Q = \mathrm{Tr}\sum_{\lambda}\sigma_{\lambda} \\
			\mathrm{s.t.}
			&\quad
			\mathrm{Tr}_A\left[
			(M_{a|x}\otimes I)O_{AB}
			\right]
			=
			\sum_{\lambda}D(a|x,\lambda)\sigma_{\lambda},
			\quad \forall a,x,\\
			&\quad
			\eta O_{AB}
			+(1-\eta)\rho_A\otimes\rho_B
			=
			\rho_{AB},\\
			&\quad \sigma_{\lambda}\succeq 0,\quad \forall \lambda .
		\end{aligned}
		\label{eq:simplified_sdp}
	\end{equation}
	If the optimal value satisfies $Q=1$, then $\rho_{AB}$ is certified to be unsteerable under all PMs~\cite{CavalcantiD2016}.

	For the $F_1$-based SVM classifiers,
	except $m=8$, any one   predicts with accuracy above $98.2\%$. For the case $m=8$, the accuracy drops to $88\%$.
	
	As the number of measurement settings $m$ increases, the method that determines steerability via SDP using only $100$ iterations of $m$ measurement directions becomes increasingly error-prone. This leads to a higher proportion of inaccurately labeled samples in the training set, which in turn degrades the prediction accuracy of the trained classifiers on strictly unsteerable states. Hence, the drop in SVM's prediction accuracy at
	$m=8$ can be attributed to the increased proportion of inaccurately labeled samples in the training set.

	\subsection{$T$-Diagonal States under Two and Three PMs}
	
	We next consider $T$-diagonal states under two and three PMs\cite{Quan2016}. A two-qubit $T$-diagonal state can be written as
	\begin{equation}
		\rho_T
		=
		\frac{1}{4}
		\left(
		I\otimes I
		+
		\sum_{i=1}^{3}t_i\sigma_i\otimes\sigma_i
		\right),
	\end{equation}
	where $T=\mathrm{diag}(t_1,t_2,t_3)$ is the correlation matrix. For two PMs, the steerability of $T$-diagonal states admits a necessary and sufficient condition~\cite{Quan2016}. Specifically,
	the state is steerable under two projective measurement settings if and only if
	\begin{equation}
		t_i^2+t_j^2>1 .
		\label{eq:Tdiag_two_meas}
	\end{equation}
	where $t_i$ and $t_j$ are the two with the largest absolute values among $t_1,t_2$ and $t_3$.
	Since this is a necessary and sufficient condition, it allows us to construct both two PMs steerable and unsteerable $T$-diagonal test subsets. For three PMs, Ref. \cite{Quan2016} presents a sufficient condition for steerability:
	\begin{equation}
		t_1^2+t_2^2+t_3^2>1 .
		\label{eq:Tdiag_three_meas}
	\end{equation}
	Because this condition is only sufficient, we use it only to construct a three PMs steerable subset.
	
\begin{figure}[!h]
		\centering
		\includegraphics[width=\columnwidth]{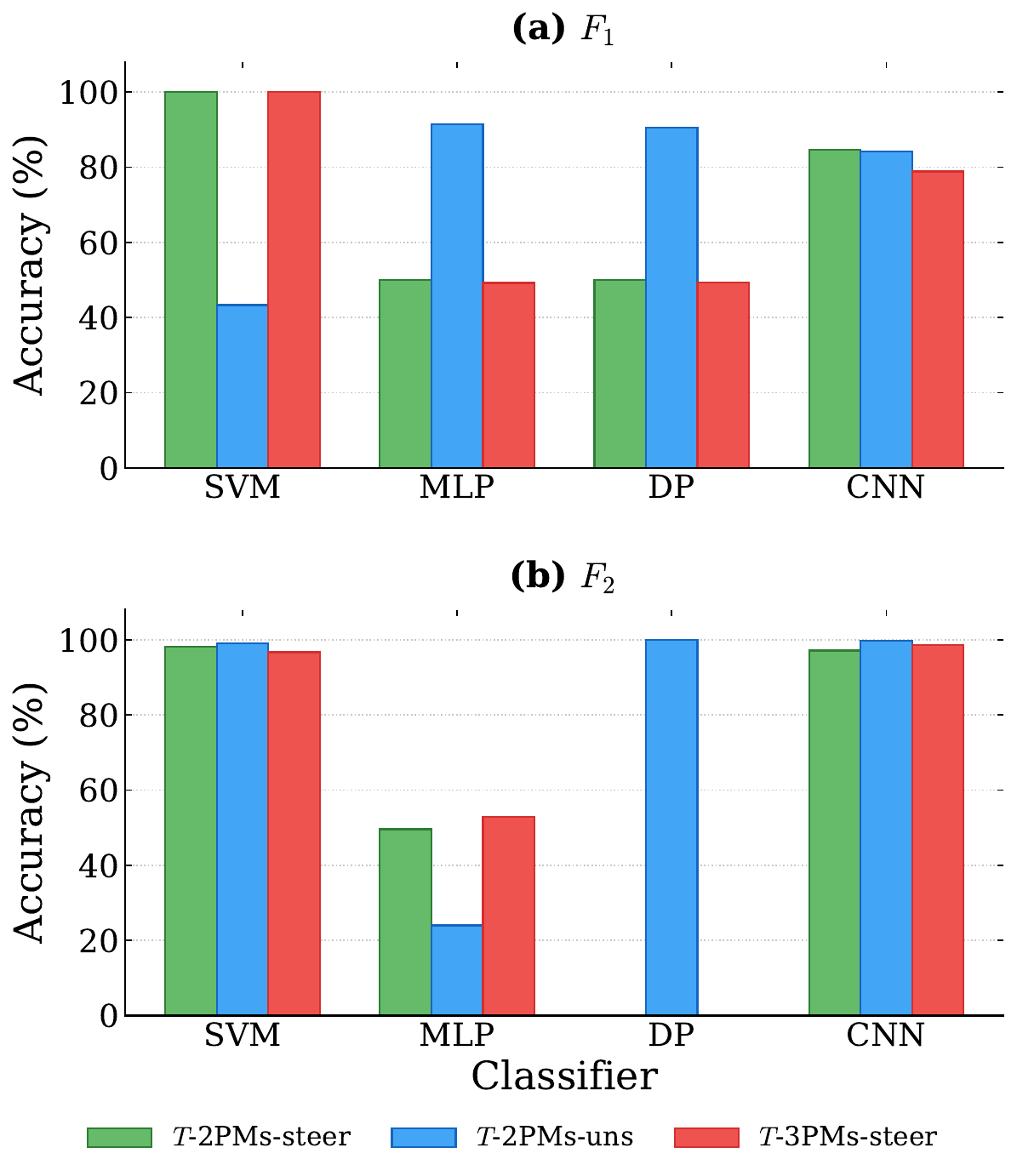}
		\caption{
			Prediction accuracies on $T$-diagonal states under two and three measurement settings.
			(a) Results based on $F_1$.
			(b) Results based on $F_2$.
			The three bars in each classifier group correspond to two-measurement steerable states, two-measurement unsteerable states, and three-measurement steerable states, respectively.
		}
		\label{fig:t_diagonal_f1_f2}
	\end{figure}

	With the above three test sets, we evaluate the generalization of SVM classifiers. The results are shown in the first three columns of Fig.~\ref{fig:t_diagonal_f1_f2} (a). The SVM-$F_1$ classifiers correctly identify
	the steerable subsets but performs poorly on the two
	PMs unsteerable subset. This result indicates a significant generalization failure on the two-PM unsteerable subset. The cause of the poor performance may also be the inaccurately labeled samples in the training set.
	
	\subsection{AVN Steerable states under 2 PMs}

	An AVN state is constructed as
	\begin{equation}
		\rho_{\rm AVN}
		=
		V|\psi(h)\rangle\langle\psi(h)|
		+
		(1-V)|\phi(h)\rangle\langle\phi(h)|,
	\end{equation}
	where
	\begin{equation}
		\begin{aligned}
			|\psi(h)\rangle
			&=
			\cos h |00\rangle+\sin h |11\rangle, \\
			|\phi(h)\rangle
			&=
			\cos h |10\rangle+\sin h |01\rangle .
		\end{aligned}
	\end{equation}
	For $h\in(0,\pi/2)$ and $V\in[0,1]\setminus\{1/2\}$, this state is entangled and can exhibit steerability  using a two-setting protocol. Therefore, in this work we employ AVN states to check the generalization of classifiers trained under two PMs.

	Unfortunately, when $m=2$, the SVM trained by the
	full-information feature $F_1$ fails to achieve good generalization on the AVN test set.
	
	\section{NEURAL NETWORK CLASSIFIERS BASED ON VECTORIZED FEATURE  $F_1$}
	
	Compared with SVMs, neural network models have more parameters that can be automatically optimized, giving them the potential to serve as robust steerability classifiers. In this section, we train an MLP with the same architecture as in Ref.~\cite{He2022detecting}, and DP using a fully connected neural-network architecture adapted from Ref.~\cite{Wang2024}.
	
	First, we shall check their performances with the cross-validation accuracy and test accuracy on random states, as well as the prediction accuracy on Werner states. For random states, DP performs comparably with SVM, while the accuracies predicted by MLP are slightly lower. For Werner states, all prediction accuracies exceed $91\%$, indicating that neural network models have better generalizability than SVM for Werner states.  Results are shown in Figure~\ref{fig:random_werner_allmodels}.
	
	Second, on the strictly unsteerable random states dataset,
	all neural network models perform comparably with SVM, with the exception of the $m=8$ case. In this particular case, DP  significantly outperforms both SVM and MLP, reaching an accuracy of  $90\%$.
	
	Third, on the three $T$-diagonal states datasets, a striking difference is that, unlike SVM, MLP and DP achieve high accuracy on the unsteerable dataset but perform no better than random guessing on the steerable one.  The results are shown in the second and the third three columns of Fig.~\ref{fig:t_diagonal_f1_f2} (a).
	
	Finally, on the steerable AVN states dataset, neither MLP nor DP achieves robust performance.
	
	\section{Extracting  canonical steering-determined feature $F_2$ and training classifiers}
	
	Since SLOCC performed on Bob's subsystem, together with local unitary transformations on the bipartite system, preserve steerability from $A$ to $B$, the canonical state $\rho_{\rm can}$ of $\rho_{AB}$ can be written as~\cite{Wang2024}
	\begin{eqnarray}
		\rho_{\rm can}
		&=&
		(U_A\otimes U_B)\tilde{\rho}_{AB}(U_A\otimes U_B)^{\dagger}
		\nonumber\\
		&=&
		\frac{1}{4}
		\left[
		I_4
		+
		(\vec{a}^{\,\prime}\cdot\vec{\sigma})\otimes I_B
		+
		\sum_{i=1}^{3}t'_{ii}\sigma_i\otimes\sigma_i
		\right],
	\end{eqnarray}
	where, for full-rank $\rho_B$,
	\begin{eqnarray}
		\tilde{\rho}_{AB}
		=
		\frac{
			(I_A\otimes\rho_B^{-1/2})\rho_{AB}(I_A\otimes\rho_B^{-1/2})
		}{
			{\rm Tr}\left[
			(I_A\otimes\rho_B^{-1/2})\rho_{AB}(I_A\otimes\rho_B^{-1/2})
			\right]
		}.
	\end{eqnarray}
	Here, $\rho_B={\rm Tr}_A(\rho_{AB})$.
	The local unitary transformations $U_X$ $(X=A,B)$ are induced by  orthogonal transformations that diagonalize the $3\times3$ correlation matrix of $\tilde{\rho}_{AB}$. Specifically, the $3\times3$ correlation matrix of $\tilde{\rho}_{AB}$
	\begin{equation}
		T_{ij}
		=
		{\rm Tr}\left[
		\tilde{\rho}_{AB}\sigma_i\otimes\sigma_j
		\right].
	\end{equation}
	is a real symmetric matrix. By the singular value decomposition theorem, there exist orthogonal matrices $O_i$ $(i=1,2)$ such that
	\begin{equation}
		O_1^{T} T O_2
		=
		{\rm diag}(t'_{11},t'_{22},t'_{33}),
	\end{equation}
	where \(t'_{11},t'_{22},t'_{33}\) are the singular values of \(T\). The local unitary transformations $U_X$ $(X=A,B)$ are obtained by
	\begin{equation}
		U_X(\vec{n}\cdot\vec{\sigma})U_X^{\dagger}
		=
		(O_i\vec{n})\cdot\vec{\sigma}.
	\end{equation}
	If $\rho_B$ is not invertible, then the rank of $\rho_B$ is one, which implies that $\rho_B$ is a pure state. It is easy to prove that in this case $\rho_{AB}$ is separable and then is unsteerable. Hence, this kind of states can be deleted from our train set or test set.
	
	The canonical state $\rho_{\rm can}$ of $\rho_{AB}$ has the same steerability from $A$ to $B$ as $\rho_{AB}$. Furthermore, in this canonical form, only six real parameters may be nonzero, and they collectively form a six-dimensional vector
	\begin{equation}
		(\vec{a}^{\,\prime},t'_{11},t'_{22},t'_{33}),
	\end{equation}
	which is just the LUTA-6 feature introduced by reference~\cite{Wang2024}. It was proved in reference~\cite{Wang2024} that the LUTA-6 feature constitutes a compact characterization of steering.  Since it is a feature derived from canonical states, it is referred to as \textit{steering-determined feature} $F_2$ in this paper.

	\subsection{Performance of Classifiers Trained by $F_2$}

	In this section, we shall train SVM, neural-network with the same MLP architecture with ~\cite{He2022detecting}, and DP using a fully connected neural-network architecture adapted from Ref.~\cite{Wang2024} for vectorized features $F_2$. Firstly, we shall check their performances with the cross-validation accuracy and test accuracy on random states, as well as the prediction accuracy on Werner states,

	\paragraph{SVM.}
	For the canonical steering-determined feature $F_2$, while it is true that SVMs can be trained to achieve high classification accuracy on random states and reasonable accuracy on Werner states, their performance on the latter nevertheless exhibits more noticeable fluctuations, with accuracy falling to $71.77\%$ for $m=2$ measurement settings. Moreover, on the strictly unsteerable random states dataset, the SVM-$F_2$ classifiers perform on par with their SVM-$F_1$ counterparts. After extracting the key feature  that determines steerability, SVM-$F_2$ classifiers achieve robust performance on any $T$-diagonal states dataset.
	Unfortunately, under the two PMs, neither the SVM trained with feature
	$F_1$ nor the one trained with feature $F_2$ generalizes to AVN states.
	
	\paragraph{MLP and DP}
	For the feature $F_2$,
	we use the same MLP architecture as that used for feature $F_1$. Although MLP-$F_1$ maintains stable performance on both random states and Werner states,
	MLP-$F_2$ shows a substantial decrease in both cross-validation and test accuracies on random states, and its Werner-state performance is also less stable. Whether on strictly unsteerable random states dataset or on the $T$-diagonal states datasets, MLP models fail to  perform stably.
	Furthermore, under the two PMs, neither the MLP trained with feature
	$F_1$ nor the one trained with feature $F_2$ generalizes to AVN states.
	
\begin{figure}[!h]
		\centering
		\includegraphics[width=\columnwidth]{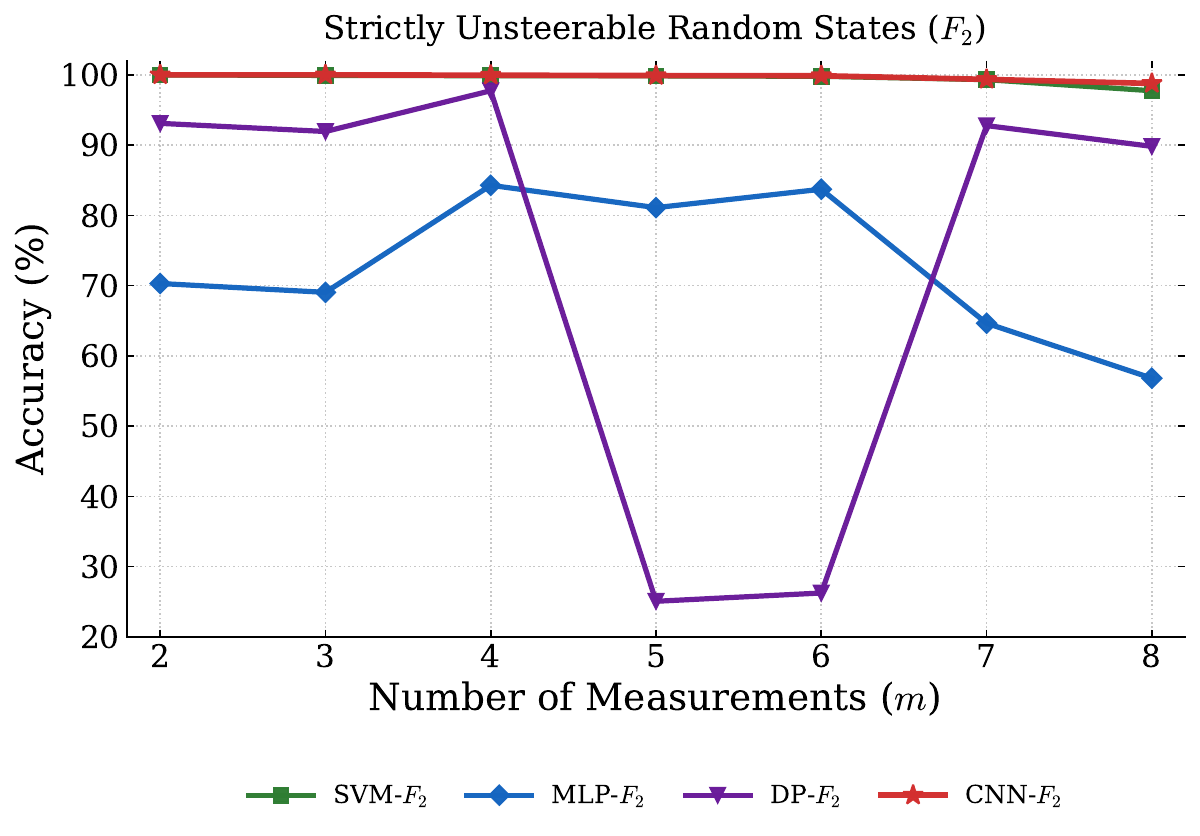}
		\caption{
			Prediction accuracies of the $F_2$-based models on strictly unsteerable random states under different measurement settings.
		}
		\label{fig:strictly_unsteerable_f2}
	\end{figure}

	For feature $F_2$, on random states and Werner states, the green, blue and purple solid lines in Figure~\ref{fig:random_werner_allmodels} represent the accuracies predicted by SVM, MLP, and DP respectively. For the $T$-diagonal states, results are displayed in the first three columns of Fig.~\ref{fig:t_diagonal_f1_f2} (b). For the strictly unsteerable random states, the green, blue and purple solid lines in Figure~\ref{fig:strictly_unsteerable_f2} show the corresponding accuracy predicted by SVM, MLP, and DP respectively.

	The failure of training robust MLP with feature $F_2$ is surprising. To further examine whether this behavior is specific to the MLP setting, we perform a DP control experiment using a fully connected neural-network architecture adapted from Ref.~\cite{Wang2024}.  Results show that  DP-$F_2$ classifiers perform as badly as MLP-$F_2$ classifiers.

	So far, two problems need to be urgently addressed. The first one is that classifiers trained on the vectorized features \(F_1\) and \(F_2\) fail to generalize well to the AVN test states. The other one is that \(F_2\), although a compact characterization of steering, is insufficient for training robust neural-network models.

	Extracting the key features that determine steerability helps to mitigate this issue. SVM trained with the feature $F_2$ performs well at $m=8$, demonstrating that extracting key features is sufficient for models like SVM. However, for neural network models, merely extracting key features is not enough to overcome the drawback of increased label inaccuracy in the training set.

	\section{Convolutional Neural Networks for Matrix-Structured Features}
	
	We suspect the one-dimensional vectorized features may hinder the training of robust steerability classifiers. In addition, the one-dimensional feature
	$F_2$ may not provide a stable input structure for fully connected neural networks. Therefore, we further consider a matrix-structure-preserving feature and train CNNs.

	\subsection{Features' Matrix Versions and CNN Architecture}
	
	Quantum steerability is an intrinsic property of quantum states, which are represented by complex-valued density matrices. To make use of the structural information encoded in these matrices, we construct matrix-structured inputs from both the original density matrix $\rho_{AB}$ and its canonical state $\rho_{\rm can}$. The former gives the matrix version of the full-information feature $F_1$, while the latter gives the matrix version of the canonical steering-related feature $F_2$.
	
	Since each quantum density matrix is complex-valued, we decompose it into two real matrices, where one matrix consists of the real components of the original complex entries, and the other consists of the imaginary components. The two real matrices of $\rho_{AB}$ and $\rho_{\rm can}$ constitute the matrix versions of feature $F_1$ and feature $F_2$ respectively. The steerability label is kept as the corresponding target label. Hence, the input of  each feature's matrix version is two real matrices, and the label is also the original label obtained by Ren. The matrix versions of features preserve the relative arrangement of the matrix elements and may train more robust steerability classifiers.
	To train CNNs, we stack
	these two components  as two channels, so that each input is  a $2\times4\times4$ multi-channel array.
	
	The CNN architecture used in this work consists of three convolutional blocks with output channel numbers $32$, $64$, and $128$, respectively. Each convolutional block uses local $3\times3$ kernels followed by a ReLU activation function. For all measurement settings $m=2,\ldots,8$, the same CNN architecture is used for both CNN-$F_1$ and CNN-$F_2$. The models are trained for $100$ epochs using the cross-entropy loss function and optimized by the Adam algorithm with a learning rate of $10^{-4}$. A four-fold cross-validation strategy is adopted, consistent with the baseline experiments.
	
	\subsection{CNN's Performance}
	
	On random states and Werner states, the classification accuracies and test accuracies of  CNN, trained by the matrix versions of feature $F_1$ and $F_2$,  are plotted in Fig.~\ref{fig:random_werner_allmodels}, together with the vectorized baseline models.
	
	On random states, both CNN-$F_1$ and CNN-$F_2$ achieve high cross-validation and test accuracies for all measurement settings. For the feature extracted from the  canonical state with the critical steerability information, the matrix version $F_2$ can train neural-network  CNN with accuracies close to or above $98\%$ in almost all cases, showing performance comparable to the strongest SVM-$F_2$ baseline and clearly outperforming MLP-$F_2$ and DP-$F_2$.
	
	For Werner-state tests, CNN-$F_1$ maintains strong generalization performance across most measurement settings.  CNN-$F_2$ also remains competitive and performs at a level comparable to SVM-$F_2$, although both $F_2$-based models exhibit certain fluctuations depending on the measurement setting. Similar to the random states, CNN-$F_2$ outperforms MLP-$F_2$ and DP-$F_2$. The accuracies predicted by CNN are plotted with red  lines in Figure~\ref{fig:random_werner_allmodels}.
	
	For the $T$-diagonal states, when choosing the full-information feature $F_1$, only when the matrix structure is preserved can  CNNs achieve  robust and  balanced performance across all the three test sets. In contrast, based on the canonical steering-related feature $F_2$,
	the key feature that determines steerability, only SVM and CNN exhibit  robust performance  among all the three test sets. Results are displayed in the last three columns of Fig.~\ref{fig:t_diagonal_f1_f2} (a) and (b).

	For strictly unsteerable states, it is also necessary to preserve the matrix structure in order to train CNNs that achieve high prediction accuracies. The red solid lines in Figure~\ref{fig:strictly_unsteerable_f2} show the corresponding accuracy predicted by CNN.
	
\begin{figure}[!h]
		\centering
		\includegraphics[width=80mm]{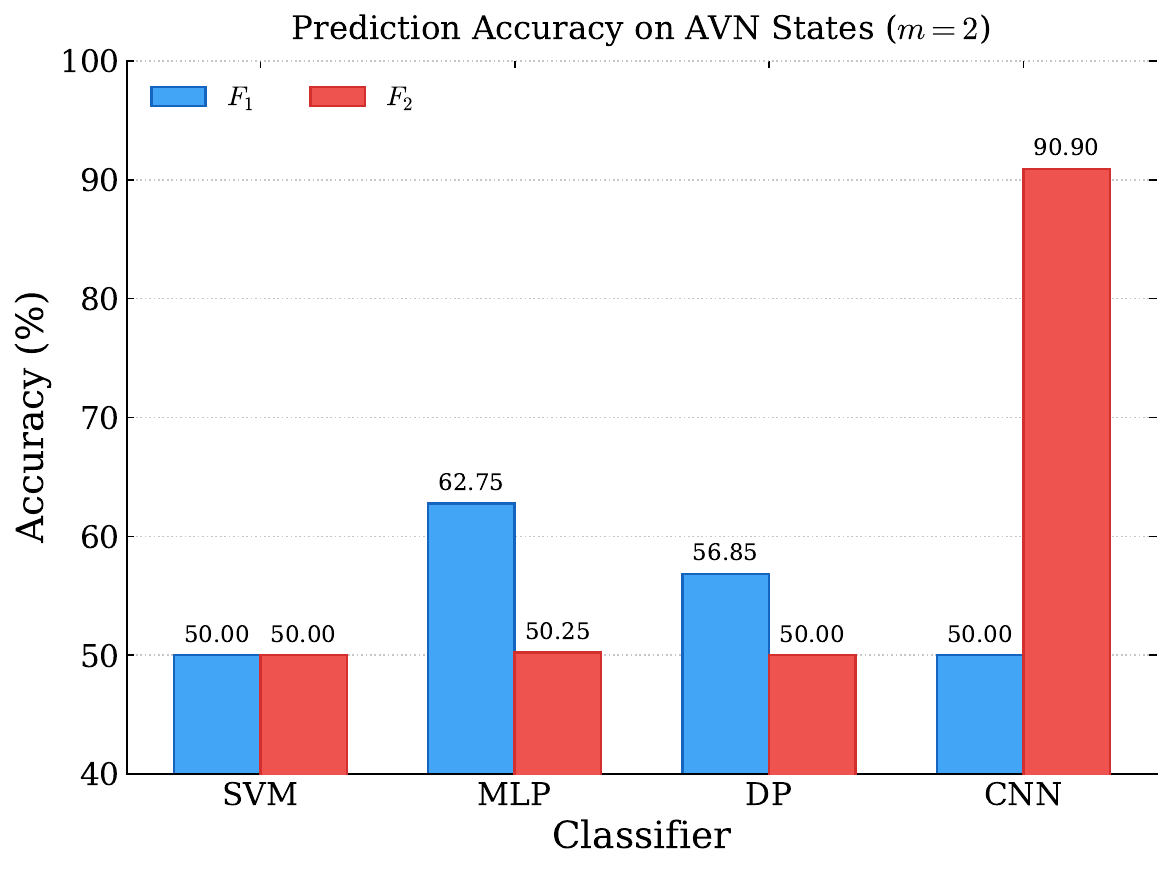}
		\caption{
			Prediction accuracies on AVN states under the two-measurement setting.
		}
		\label{fig:avn_m2}
	\end{figure}

	More encouragingly, although CNN-$F_1$ cannot generalize to the prediction of AVN states' steerability, CNN-$F_2$ achieves a prediction accuracy of $90.9\%$. Results are shown in Figure~\ref{fig:avn_m2}.Thus, CNN-\(F_2\) resolves the two issues raised above.
	
	These results indicate that both the matrix-structure-preserving and the extracting of governing steerability
	critical features lead to the matrix version feature $F_2$, which enables the training of CNNs  with the most robust steerability classification capacity.

	\section{Prediction of Steerability Boundaries for Axially Symmetric States}
	
	Thus far, $F_2$-based CNNs exhibit the greatest robustness among all steerability classifiers.
	In this section, we apply them to predict the steerability boundaries of axially symmetric states. These states,   a special subclass of $T$-diagonal states~\cite{Nguyen2020}, can be written as
	\begin{equation}
		\rho_{\rm ax}(s,t)
		=
		\frac{1}{4}
		\left(
		I\otimes I
		+
		s\sigma_x\otimes\sigma_x
		+
		s\sigma_y\otimes\sigma_y
		+
		t\sigma_z\otimes\sigma_z
		\right),
	\end{equation}
	i.e., the correlation matrix $T$ with the form \begin{equation}
		T=\mathrm{diag}(s,s,t).
	\end{equation}
	The positivity of $\rho_{\rm ax}(s,t)$ imposes the physical constraint
	\begin{equation}
		-1\leq t \leq 1-2|s| ,
	\end{equation}
	so the admissible states occupy a triangular region in the $(s,t)$ plane.
	By specializing the necessary and sufficient condition for 2 PMs in Eq.~\eqref{eq:Tdiag_two_meas} to the case of $T=\mathrm{diag}(s,s,t)$, the steering boundary  is given by
	\begin{equation}
		s^2+\max(s^2,t^2)=1 .
	\end{equation}
	Similarly, the sufficient condition for 3 PMs in Eq.~\eqref{eq:Tdiag_three_meas} gives the sufficient boundary
	\begin{equation}
		2s^2+t^2=1 .
	\end{equation}
	\begin{figure}[!h]
		\centering
		\includegraphics[width=\columnwidth]{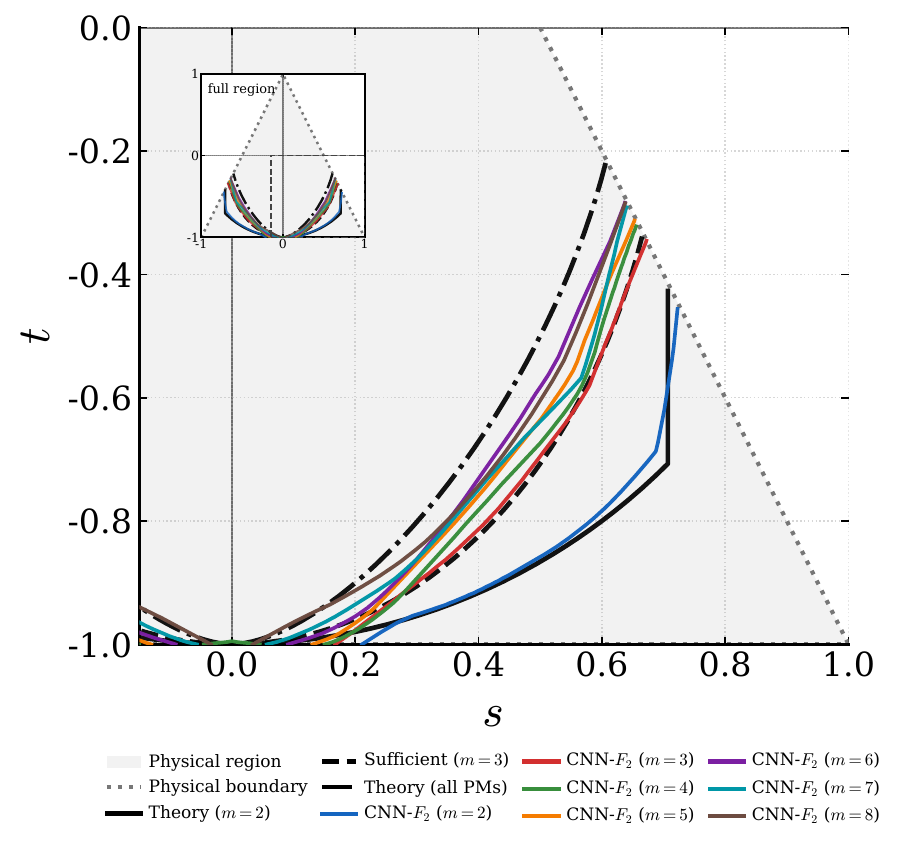}
		\caption{
			CNN-$F_2$ prediction of steering boundaries for axially symmetric states.
		}
		\label{fig:axial_boundary_cnn_f2}
	\end{figure}
	
	The steerability boundary predicted by CNN-$F_2$  for $m=3$ is close to that from the  sufficient condition by (25), suggesting that its performance is comparable to the three PMs' sufficient criterion. For $m=2$, the predicted boundary also follows the overall trend of the necessary and sufficient theoretical boundary, although visible deviations remain in part of the physical region. Furthermore, we also predict the steerability boundary of axially symmetric states under $m$ PMs for $m=4,5,6,7,8$. The predicted boundaries are shown in Fig. ~\ref{fig:axial_boundary_cnn_f2}. To further verify these predictions, we use SDP to identify specific examples that are steerable under the corresponding \(m\) PMs; these examples are listed in Appendix.C In other words, CNN-$F_2$ can predict how many projective measurements are needed to detect the steerability of steerable axially symmetric states.

	\section{Conclusion}
	
	In this work, we investigated the generalization ability of quantum steering classifiers from the perspective of feature representation. We showed that extracting steering-determined features alone is insufficient for training robust neural-network classifiers, while preserving the matrix structure of the canonical state enables CNN-\(F_2\) to achieve stable performance across different test sets. As an application, we used the robust CNN-\(F_2\) classifiers to predict the steering boundaries of axially symmetric states and to estimate the number of projective measurements required for steering detection. These results suggest that matrix-structure-preserving feature learning may provide a useful strategy for future data-driven studies of steerability in higher-dimensional or multipartite quantum systems.

	\begin{acknowledgments}
    \section*{ACKNOWLEDGMENTS}
	Thanks to Dr. Su Han for the constructive discussion. H. X. Meng was supported by the Beijing Natural Science Foundation (Grant No. 1262039), and the National Natural Science Foundations of China(Grant No. 12461087). Z. Y. Li was  supported by the National Natural Science Foundations of China (Grants No. 12571560).

	\end{acknowledgments}

	\onecolumngrid
	\appendix
	
    \section{Detailed Accuracy Tables}
    \label{app:accuracy}
    
    All numerical values are given in percentages. 
    
    \begin{table}[H]
    	\caption{
    		Accuracies of CNN models on random states, Werner states, and strictly unsteerable random states.
    		\label{tab:supp_cnn_accuracy}
    	}
    	\begin{ruledtabular}
    		\begin{tabular}{llccccccc}
    			Feature & Test & $m=2$ & $m=3$ & $m=4$ & $m=5$ & $m=6$ & $m=7$ & $m=8$ \\
    			$F_1$ & Crossrate & 98.00 & 97.80 & 97.10 & 96.50 & 96.70 & 95.40 & 97.90 \\
    			$F_1$ & Testrate  & 98.70 & 97.50 & 97.20 & 96.25 & 96.20 & 94.20 & 98.20 \\
    			$F_1$ & Werner    & 97.01 & 98.96 & 95.42 & 97.16 & 98.18 & 91.82 & 93.58 \\
    			$F_1$ & Strictly unsteerable & 99.98 & 99.86 & 99.98 & 99.58 & 99.86 & 98.72 & 88.02 \\
    			$F_2$ & Crossrate & 99.10 & 99.40 & 99.35 & 99.45 & 99.30 & 98.00 & 99.30 \\
    			$F_2$ & Testrate  & 99.60 & 99.45 & 99.50 & 99.10 & 99.45 & 98.90 & 99.80 \\
    			$F_2$ & Werner    & 80.77 & 92.02 & 92.13 & 94.80 & 96.81 & 84.40 & 95.52 \\
    			$F_2$ & Strictly unsteerable & 100.00 & 100.00 & 99.96 & 99.90 & 99.88 & 99.36 & 98.78 \\
    		\end{tabular}
    	\end{ruledtabular}
    \end{table}

    \begin{table}[H]
    	\caption{
    		Accuracies of SVM models on random states, Werner states, and strictly unsteerable random states.
    		\label{tab:supp_svm_accuracy}
    	}
    	\begin{ruledtabular}
    		\begin{tabular}{llccccccc}
    			Feature & Test & $m=2$ & $m=3$ & $m=4$ & $m=5$ & $m=6$ & $m=7$ & $m=8$ \\
    			$F_1$ & Crossrate & 98.43 & 97.39 & 97.02 & 95.91 & 96.72 & 95.04 & 98.22 \\
    			$F_1$ & Testrate  & 97.90 & 97.65 & 96.65 & 96.55 & 97.50 & 95.35 & 98.35 \\
    			$F_1$ & Werner    & 91.46 & 83.36 & 97.67 & 97.81 & 99.86 & 92.56 & 96.17 \\
    			$F_1$ & Strictly unsteerable & 100.00 & 100.00 & 99.84 & 99.90 & 99.76 & 98.96 & 88.36 \\
    			$F_2$ & Crossrate & 99.20 & 97.39 & 99.50 & 99.20 & 98.72 & 98.04 & 99.22 \\
    			$F_2$ & Testrate  & 99.25 & 99.50 & 99.65 & 99.55 & 99.60 & 98.40 & 99.70 \\
    			$F_2$ & Werner    & 71.77 & 93.19 & 84.00 & 97.29 & 98.12 & 88.70 & 85.56 \\
    			$F_2$ & Strictly unsteerable & 100.00 & 99.92 & 99.88 & 99.86 & 99.80 & 99.32 & 97.74 \\
    		\end{tabular}
    	\end{ruledtabular}
    \end{table}

    \begin{table}[H]
    	\caption{
    		Accuracies of MLP and DP models on random states, Werner states, and strictly unsteerable random states.
    		\label{tab:supp_mlp_dp_accuracy}
    	}
    	\begin{ruledtabular}
    		\begin{tabular}{lllccccccc}
    			Model & Feature & Test & $m=2$ & $m=3$ & $m=4$ & $m=5$ & $m=6$ & $m=7$ & $m=8$ \\
    			MLP & $F_1$ & Crossrate & 95.70 & 94.75 & 92.10 & 91.24 & 92.45 & 91.20 & 96.45 \\
    			MLP & $F_1$ & Testrate  & 94.50 & 94.60 & 92.34 & 90.35 & 92.48 & 92.05 & 96.54 \\
    			MLP & $F_1$ & Werner    & 97.45 & 94.95 & 99.91 & 96.63 & 94.97 & 99.53 & 97.88 \\
    			MLP & $F_1$ & Strictly unsteerable & 100.00 & 99.94 & 99.96 & 99.96 & 99.98 & 99.04 & 73.02 \\
    			MLP & $F_2$ & Crossrate & 61.85 & 62.25 & 61.45 & 61.24 & 67.15 & 60.92 & 66.05 \\
    			MLP & $F_2$ & Testrate  & 61.92 & 62.13 & 63.42 & 61.70 & 67.05 & 61.05 & 66.54 \\
    			MLP & $F_2$ & Werner    & 64.30 & 64.30 & 81.87 & 79.93 & 54.59 & 64.30 & 70.92 \\
    			MLP & $F_2$ & Strictly unsteerable & 70.30 & 69.04 & 84.28 & 81.10 & 83.72 & 64.64 & 56.82 \\
    			\noalign{\vskip 2pt}
    			\colrule
    			\noalign{\vskip 2pt}
    			DP  & $F_1$ & Crossrate & 98.30 & 97.30 & 97.20 & 95.70 & 96.60 & 94.70 & 98.00 \\
    			DP  & $F_1$ & Testrate  & 98.60 & 97.27 & 97.15 & 95.65 & 96.54 & 94.60 & 97.85 \\
    			DP  & $F_1$ & Werner    & 99.35 & 91.51 & 99.19 & 97.89 & 97.78 & 99.58 & 99.99 \\
    			DP  & $F_1$ & Strictly unsteerable & 100.00 & 99.98 & 99.84 & 99.70 & 99.50 & 98.22 & 90.10 \\
    			DP  & $F_2$ & Crossrate & 59.45 & 58.75 & 54.40 & 55.60 & 56.90 & 59.05 & 58.80 \\
    			DP  & $F_2$ & Testrate  & 58.56 & 59.01 & 54.25 & 55.48 & 56.85 & 59.00 & 58.65 \\
    			DP  & $F_2$ & Werner    & 50.00 & 50.00 & 50.00 & 50.00 & 50.00 & 50.00 & 50.00 \\
    			DP  & $F_2$ & Strictly unsteerable & 93.10 & 91.94 & 97.74 & 25.06 & 26.24 & 92.78 & 89.82 \\
    		\end{tabular}
    	\end{ruledtabular}
    \end{table}

    \begin{table}[H]
    	\caption{
    		Measurement-specific accuracies on AVN states and $T$-diagonal states.
    		\label{tab:supp_measurement_specific}
    	}
    	\begin{ruledtabular}
    		\begin{tabular}{lcccc}
    			Model 
    			& AVN, $m=2$
    			& $T$-diagonal steerable, $m=2$
    			& $T$-diagonal unsteerable, $m=2$
    			& $T$-diagonal steerable, $m=3$ \\
    			CNN-$F_1$ & 50.00 & 84.62 & 84.22 & 78.92 \\
    			CNN-$F_2$ & 90.90 & 97.16 & 99.80 & 98.66 \\
    			SVM-$F_1$ & 50.00 & 100.00 & 43.32 & 100.00 \\
    			SVM-$F_2$ & 50.00 & 98.18 & 99.10 & 96.72 \\
    			MLP-$F_1$ & 62.75 & 49.96 & 91.48 & 49.24 \\
    			MLP-$F_2$ & 50.25 & 49.62 & 24.04 & 52.92 \\
    			DP-$F_1$ & 56.85 & 49.98 & 90.52 & 49.30 \\
    			DP-$F_2$ & 50.00 & 0.00 & 100.00 & 0.00 \\
    		\end{tabular}
    	\end{ruledtabular}
    \end{table}
    \section{Optimal hyperparameters of SVM classifiers}
    \label{app:svm_parameters}
    
    The optimal hyperparameters of the RBF-kernel SVM classifiers were selected by five-fold cross-validation over the exponential grid
    \[
    C=2^c,\qquad \gamma=2^g .
    \]
    The selected values for SVM-F1 and SVM-F2 are listed in Tables~\ref{tab:svm_f1_parameters} and~\ref{tab:svm_f2_parameters}, respectively.
    
    \begin{table}[htbp]
    	\centering
    	\caption{
    		Optimal hyperparameters of the RBF-kernel SVM classifier using feature F1.
    	}
    	\label{tab:svm_f1_parameters}
    	\small
    	\renewcommand{\arraystretch}{1.15}
    	\setlength{\tabcolsep}{7pt}
    	\begin{tabular}{lccccccc}
    		\toprule
    		Hyperparameter & $m=2$ & $m=3$ & $m=4$ & $m=5$ & $m=6$ & $m=7$ & $m=8$ \\
    		\midrule
    		$C$      & $2^{6}$  & $2^{6}$  & $2^{5}$  & $2^{5}$  & $2^{6}$  & $2^{5}$  & $2^{5}$  \\
    		$\gamma$ & $2^{-5}$ & $2^{-5}$ & $2^{-5}$ & $2^{-5}$ & $2^{-5}$ & $2^{-5}$ & $2^{-5}$ \\
    		\bottomrule
    	\end{tabular}
    \end{table}
    
    \begin{table}[htbp]
    	\centering
    	\caption{
    		Optimal hyperparameters of the RBF-kernel SVM classifier using feature F2.
    	}
    	\label{tab:svm_f2_parameters}
    	\small
    	\renewcommand{\arraystretch}{1.15}
    	\setlength{\tabcolsep}{7pt}
    	\begin{tabular}{lccccccc}
    		\toprule
    		Hyperparameter & $m=2$ & $m=3$ & $m=4$ & $m=5$ & $m=6$ & $m=7$ & $m=8$ \\
    		\midrule
    		$C$      & $2^{5}$  & $2^{5}$  & $2^{2}$  & $2^{5}$  & $2^{6}$  & $2^{6}$  & $2^{5}$  \\
    		$\gamma$ & $2^{-3}$ & $2^{-2}$ & $2^{-1}$ & $2^{-3}$ & $2^{-5}$ & $2^{-3}$ & $2^{-1}$ \\
    		\bottomrule
    	\end{tabular}
    \end{table}

    \section{SDP-verified axially symmetric states and measurement directions}
    \label{app:axial_examples}

    In this appendix, for each fixed number of projective measurements
    $m=3,4,5,6,7,8$, we list a representative steerable axially symmetric
    state predicted by CNN-$F_2$ to be detectable with $m$ projective
    measurements. For each listed state, we further provide one set of
    $m$ projective measurement directions. These measurement settings are
    then used in the SDP test to verify that the corresponding state is
    indeed steerable under the $m$-measurement setting.
    
    The measurement directions are denoted by
    $\vec{n}_x=(n_{x,1},n_{x,2},n_{x,3})$, $x=1,\ldots,m$, and define the
    two-outcome projective measurements
    \begin{equation}
    	P_{\pm|x}=\frac{1}{2}(I\pm \vec{n}_x\cdot \vec{\sigma}).
    \end{equation}
    The listed measurement settings therefore provide explicit
    SDP-verified examples supporting the CNN-$F_2$ predictions for
    axially symmetric states.
    
    \subsection{The case of \(m=3\)}
    
    For three projective measurements, a representative SDP-verified steerable axially symmetric state is given by
    \begin{equation}
    	(s,t)=(0.671436,-0.617085).
    \end{equation}
    One set of measurement directions detecting its steerability is
    \begin{equation}
    	\begin{aligned}
    		\vec n_1 &= (0.9430,-0.2200,-0.2497),\\
    		\vec n_2 &= (-0.1596,0.1196,-0.9799),\\
    		\vec n_3 &= (-0.4206,-0.9072,-0.0012).
    	\end{aligned}
    \end{equation}
    
    \subsection{The case of \(m=4\)}
    
    For four projective measurements, a representative SDP-verified steerable axially symmetric state is given by
    \begin{equation}
    	(s,t)=(0.655490,-0.370736).
    \end{equation}
    One set of measurement directions detecting its steerability is
    \begin{equation}
    	\begin{aligned}
    		\vec n_1 &= (0.8945,-0.3502,-0.2779),&
    		\vec n_2 &= (0.5208,0.8449,-0.1223),\\
    		\vec n_3 &= (-0.0485,0.9893,0.1376),&
    		\vec n_4 &= (0.7556,-0.4164,0.5057).
    	\end{aligned}
    \end{equation}
    
    \subsection{The case of \(m=5\)}
    
    For five projective measurements, a representative SDP-verified steerable axially symmetric state is given by
    \begin{equation}
    	(s,t)=(0.577012,-0.542849).
    \end{equation}
    One set of measurement directions detecting its steerability is
    \begin{equation}
    	\begin{aligned}
    		\vec n_1 &= (0.4096,0.0868,-0.9081),&
    		\vec n_2 &= (0.9014,0.2681,-0.3400),\\
    		\vec n_3 &= (0.7158,-0.5757,-0.3951),&
    		\vec n_4 &= (0.3880,0.7555,0.5279),\\
    		\vec n_5 &= (0.5032,-0.7906,0.3488).
    	\end{aligned}
    \end{equation}
    
    \subsection{The case of \(m=6\)}
    
    For six projective measurements, a representative SDP-verified steerable axially symmetric state is given by
    \begin{equation}
    	(s,t)=(0.559475,-0.523110).
    \end{equation}
    One set of measurement directions detecting its steerability is
    \begin{equation}
    	\begin{aligned}
    		\vec n_1 &= (-0.5595,0.1135,-0.8210),&
    		\vec n_2 &= (-0.2277,-0.2440,0.9427),\\
    		\vec n_3 &= (-0.1745,-0.9437,-0.2809),&
    		\vec n_4 &= (-0.9927,0.0768,-0.0927),\\
    		\vec n_5 &= (0.5636,-0.7768,0.2809),&
    		\vec n_6 &= (-0.2854,-0.9045,0.3168).
    	\end{aligned}
    \end{equation}
    
    \subsection{The case of \(m=7\)}
    
    For seven projective measurements, a representative SDP-verified steerable axially symmetric state is given by
    \begin{equation}
    	(s,t)=(0.199606,-0.950116).
    \end{equation}
    One set of measurement directions detecting its steerability is
    \begin{equation}
    	\begin{aligned}
    		\vec n_1 &= (0.5459,-0.1284,0.8280),&
    		\vec n_2 &= (-0.3335,-0.3498,0.8755),\\
    		\vec n_3 &= (-0.1154,-0.1394,-0.9835),&
    		\vec n_4 &= (0.4332,0.7497,0.5003),\\
    		\vec n_5 &= (0.7241,-0.6212,-0.2995),&
    		\vec n_6 &= (-0.8378,0.4943,-0.2320),\\
    		\vec n_7 &= (0.3942,-0.5001,-0.7710).
    	\end{aligned}
    \end{equation}
    
    \subsection{The case of \(m=8\)}
    
    For eight projective measurements, a representative SDP-verified steerable axially symmetric state is given by
    \begin{equation}
    	(s,t)=(0.567935,-0.541529).
    \end{equation}
    One set of measurement directions detecting its steerability is
    \begin{equation}
    	\begin{aligned}
    		\vec n_1 &= (-0.1972,-0.9801,0.0222),&
    		\vec n_2 &= (0.8116,-0.4892,-0.3194),\\
    		\vec n_3 &= (-0.7993,-0.3720,0.4719),&
    		\vec n_4 &= (0.4944,-0.4879,0.7194),\\
    		\vec n_5 &= (-0.3942,0.1657,0.9040),&
    		\vec n_6 &= (0.8181,-0.3558,-0.4517),\\
    		\vec n_7 &= (-0.6330,0.5123,0.5804),&
    		\vec n_8 &= (-0.6176,-0.4170,-0.6668).
    	\end{aligned}
    \end{equation}

\end{document}